\documentclass[preprint,showpacs,preprintnumbers,amsmath,amssymb]{revtex4}
\usepackage{graphicx}

\usepackage{dcolumn}

\usepackage{bm}

\begin{document}

\title{Magnetically driven ferroelectric order in Ni$_3$V$_2$O$_8$}
\author{G. Lawes,$^{1,+}$ A. B. Harris,$^{2}$ T. Kimura,$^1$N. Rogado,$^{3,*}$ R. J.  Cava,$^3$ A. Aharony,$^4$ O. Entin-Wohlman,$^4$\\T.  Yildirim,$^5$ M. Kenzelmann,$^{5,6}$ C. Broholm,$^{5,6}$,and A. P. Ramirez$^{1,7}$}
\affiliation {$^1$ Los Alamos National Laboratory, Los Alamos, New
Mexico 87545}
\affiliation{$^2$Department of Physics and Astronomy, University
of Pennsylvania, Philadelphia, PA, 19104}
\affiliation{$^3$Department of Chemistry and Princeton Materials
Institute, Princeton University, Princeton, New Jersey 08544}
\affiliation{$^4$School of Physics and Astronomy, Raymond and
Beverly Sackler Faculty of Exact Sciences,\\ Tel Aviv University,Tel Aviv 69978, Israel}
\affiliation{$^5$NIST Center for Neutron Research, Gaithersburg, MD 20899}
\affiliation{$^6$Department of Physics and Astronomy, Johns
Hopkins University, Baltimore, MD 21218}
\affiliation{$^7$Bell Labs, Lucent Technologies, 600 Mountain
Avenue, Murray Hill, NJ 07974}

\date{\today}

\begin{abstract}
We show that for Ni$_3$V$_2$O$_8$ long-range ferroelectric and
incommensurate magnetic order appear simultaneously in
a single phase transition.  The temperature and magnetic field dependence
of the spontaneous polarization show a strong coupling between magnetic and ferroelectric orders.  We determine the magnetic symmetry of this system by constraining the
data to be consistent with Landau theory for continuous
phase transitions.  This phenomenological theory explains
our observation the spontaneous polarization is restricted to lie along
the crystal \underline b axis and predicts that the magnitude should be proportional
to a magnetic order parameter.
\end{abstract}


\maketitle

The coupling between long-range magnetic and ferroelectric order has been studied since the 1960s.\cite{BIRSS,smolchup,SCHMID} Although a number of systems which  are ferroelectric at high
temperatures become magnetically ordered at a lower temperature,\cite{smolchup,LOTTER}the simultaneous appearance of both kinds of order at a single
phase transition is much less common. For the most part, studies of these
multiferroic materials
have focussed on {\it commensurate} magnets.  The magnetoelectric properties
of these systems are typically discussed in terms of magnetic symmetry groups. The onset of magnetic ordering requires breaking time-reversal symmetry, while ferroelectric order breaks spatial inversion symmetry.  Therefore, only magnetic groups having the proper symmetries allow the
possibility of simultaneous magnetic and ferroelectric order.

Recent studies of systems underoing simultaneous ferroelectric order at a magnetic transition have identified new multiferroic compounds, including the perovskite Rare Earth manganites TbMnO$_3$and DyMnO$_3$\cite{tkimura}, and TbMn$_2$O$_5$\cite{NATURE}.  However, the magnetic structures of these
materials are complex\cite{KAJI,chapon,kobayashi}, which makes an investigation of their magnetoelectric properties based on symmetry analysis problematic. In contrast, the magnetic structure of Ni$_3$V$_2$O$_8$ (NVO),which we have identified as a multiferroic material, is better determined.  There are extensive neutron data on this compound \cite{NVO}, and  a symmetry analysis of the data (discussed below) constrains the symmetries which are consistent with a continuous transition.  Here we take
advantage of our analysis of the magnetic symmetry of NVO to develop a phenomenological Landau theory which provides an explanation of our observations of ferroelectric order induced by magnetic ordering.  This provides an alternate route to understanding the magnetoelectric coupling in multiferroic
materials beyond the traditional magnetic symmetry group analysis.

We briefly review earlier results\cite{ROGADO,NVO,CALVO} for the structure (magnetic and crystal) of (NVO) since its symmetry is crucial to the development of our model. NVO is a magnetic insulator consisting of planes of spin-1 Ni$^{2+}$ ions arranged in a Kagom\'e staircase lattice. Fig. 1a shows the positions of the two kinds of Ni$^{+2}$ spins which we call ``spine'' and ``cross tie'' spins.  Competition between several weak magnetic interactions and anisotropies yields the complex phase diagram of Fig. 2 and the variety of magnetic structures illustrated in Fig 1b-d.  Cooling at ${\bf H}=0$, one first enters  a high temperature incommensurate\cite{IC}(HTI) phase at $T_H=9.1$K (Fig. 1b), then a low-temperature incommensurate (LTI) phase at $T_L=6.3$K (Fig. 1c).Below 3.9~K the system displays two slightly different canted antiferromagnet (CAF) phases (Fig. 1d). The transitions involving the HTI phase are continuous, whereas that from the LTI to one of the CAF phases is discontinuous. In the HTI phase the long-range order is mostly on the ``spine" sites with their spins parallel to {\bf a}, while in the LTI phase the spine and cross tie spins rotate within an {\bf a-b} plane as shown in Fig. 1b-c. We used representation analysis\cite{WILLS} to determine the symmetry of these phases rather than attempting to acquire the vast amount of data which would be necessary to experimentally fix the several complex order parameters[ {\it e. g.} see Eq. (2), below] for these magnetic structures. Since the irreducible representation associated with magnetic ordering are one dimensional, each magnetic order parameter is characterized by a phase factor of unit magnitude which specifies how the order parameter transforms under the symmetry operations which leave the incommensurate wavevector invariant.  Thus the HTI order parameter, $\sigma_{\rm H}$ is odd under a two-fold rotation about the \underline a-axis and is even under the glide plane whose mirror plane is an \underline a-\underline b plane. In the LTI phase an additional order parameter, $\sigma_{\rm L}$ appears which is even under both operations.

  Single crystals of NVO were grown from BaO-V$_2$O$_5$ flux\cite{NVO}. The single crystal samples were oriented using a Laue X-ray diffractometer. To probe the ferroelectric order,we sputtered two gold electrodes on opposite faces of the crystals and measured both the pyroelectric current at fixed ${\bf H}$, and the magnetoelectric current at fixed $T$, using a Keithely electrometer.  We then integrated the current to find the spontaneous polarization ${\bf P}$, corresponding to the ordered moment arising from the ferroelectric state.  We aligned the ferroelectric domains by applying a ($\sim 2$ kV/cm) polarizing electric field, $E_0$, as the sample was cooled throught he transition temperature. After removing the electric field, we measured the polarization $P_a$, $P_b$, and $P_c$, along each of the three crystal axes.  To within experimental error, only $P_b$ was found to be nonzero.
  
   Figures 3a and 3b show $P_b$, the electric polarization along the \textbf{b} axis, versus $T$ at zero applied voltage for fixed magnetic fields applied along the \textbf{a} and \textbf{c} axes, respectively. Since the sign of the polarization in zero electric field changed when the sign of the polarizing field$E_0$ was changed, we infer that this polarization arises from ferroelectric order in NVO.  From data for the magnetic field along different crystal axes, it became clear that the region in which $P_b$ was nonzero coincided exactly with the region in which the LTI phase existed, as shown in Figure 2.  Furthermore the hysteresis in ferroelectric order as a function of magnetic field or temperature is connected with the fact thatthe LTI-CAF phase transition is discontinuous.
   
   The isothermal data shown in Fig. 3b and 3d, corroborate the above picture.  Fig. 3d shows $P_b$  versus the magnetic field along \textbf{c}, $H_c$,  at $T=5$K. Atlow $H_c$, $P_b$ is insensitive to the external magnetic field. As$H_c$ is increased, the sample undergoes CAF ordering, which completely suppresses the spontaneous polarization.  On decreasing$H_c$,  $P_b$ returns to the initial value. The field-hysteresis is attributed to the first order LTI to CAF transition.  Fig. 3b shows $P_b$ at $T=$2K, versus $\bf H||a$. At small $H_a$,there is no ferroelectric order in the CAF phase. Increasing $H_a$ however produces a spontaneous polarization in the LTI phase which is independent of the sign of the magnetic field.  The crucial observation is that we can magnetically gate NVO to either suppress or promote ferroelectric order, depending only on the direction of the applied field and the temperature.

We now give a phenomenological explanation of our results. In the HTI phase the existence of incommensurate magnetic order does not induce a nonzero value of ${\bf P}$ in the HTI phase because the HTI phase is inversion symmetric. This symmetry is not easy to establish directly by analyzing the neutron diffraction data.  If one assumes the spin amplitudes are restricted according to a representation analysis,\cite{WILLS} one still has to determine the complex-valued order parameters $\sigma_{H;i}(k)=\sigma_{H;i}(-k)^*$ which characterize the HTI phase, where $k$ denotes the wavevector  of the incommensurate ordering and $i=1,...,6$, corresponds to the six spin components (the \textbf{a}, \textbf{b}, and \textbf{c} components of the spine and cross tie spins).  In the LTI phase the additional order parameters $\sigma_{L;i}(k)=\sigma_{L;i}(-k)^*$ appear. Representation analysis is based on the so-called ``little group,"$G_k$, which leaves the wavevector invariant.  But this analysis does not take into account any additional symmetries (such as spatial inversion ${\cal I}$) which the crystal possesses.  We use the inversion symmetry of the crystal lattice to define the order parameters so that they obey
\begin{eqnarray}{\cal I} 
\sigma_{X;i}(k)= \sigma_{X;i}(k)^* \ ,
\label{IEQ} 
\end{eqnarray}
where $X$ denotes either H or L. The quadratic terms in the Landau free energy which describe the appearance of the HTI phase are
\begin{eqnarray}
F &=& \sum_{i,j} F_{ij} \sigma_{H;i}(-k) \sigma_{H;j}(k) \ ,
\label{SYMEQ} 
\end{eqnarray}
where $F_{i,j}=F_{j,i}^*$.  Then Eq. (\ref{IEQ}) implies that the coefficients $F_{ij}$ are all real valued. Since the spin components are those of the critical (i. e.the one whose eigenvalue becomes negative as $T$ is lowered through $T_{\rm H}$) eigenvectorof the real symmetric matrix $F$, the amplitudes $\sigma_{H,i}$ are all real valued, apart from a common overall phase:
\begin{eqnarray}
\sigma_{H;i}(k) = c_i e^{i \phi_H} \sigma_H \ , \label{EQ2}
\end{eqnarray}

where the $c_i$, obtained from the critical eigenvector, are real. Thus the HTI phase is described by a single complex valued order parameter $e^{i\phi_{\rm H}}\sigma_{\rm H}$,whose overall phase, $\phi_{\rm H}$, is not fixed at this level of analysis. The overall phase $\phi_{\rm H}$ can be eliminated by redefining the origin of coordinates. The fact that the resulting Fourier components $\sigma_{H;i}(k)$ are all real means that the magnetic structure is invariant under spatial inversion.  Because the relative phases of the spin components are difficult to determine from an analysis of the diffraction data, this formal argument is essential to establish that the HTI phase is inversion symmetric and therefore that the magnetic ordering can not induce a spontaneous electric polarization.  Since the CAF phaseis also inversion symmetric (Fig. 1d), ferroelectricity is not induced by magnetic order in that phase either.  These results confirm that this model based on our extension of representation analysis successfully predicts the experimental observation that there is no ferroelectric order in the HTI or CAF phases.  We can also understand why when a sufficiently large magnetic field is applied so as to enter the CAF phase (Fig. 2b), the spontaneous polarization abruptly disappears. 

We now introduce a phenomenological model to explain the symmetry of the magnetoelectric  coupling which induces the spontaneous polarization. In principle, this symmetry might be deduced from a study of magnetic symmetry,\cite{BIRSS} but we have found it much simpler to directly obtain these results using Landau theory combined with representation analysis. While the specific details of this discussion apply to NVO, the general approach is applicable to other systems. To describe the incommensurate phases and ferroelectric behavior in a single model, we write the Landau free energy as

\begin{eqnarray}
F &=& a (T-T_{\rm H}) \sigma_{\rm H}^2 + b (T-T_{\rm L}) \sigma_{\rm L}^2\nonumber\\ && \ + {\cal O} (\sigma^4)+ (2 \chi_E)^{-1} {\bf P}^2 + V \ ,\label{FEQ}
\end{eqnarray}

where $a$, and $b$ are constants and $\chi_E$ is the electric susceptibility.  (We assume that the wavevectors of the two incommensurate phase are locked to be equal.) In a conventional ferroelectric ${\bf P}$ becomes nonzero when $\chi_E$ becomes infinite, often associated with a structural phase transition. Here $\chi_E$ is finite and the appearance of a nonzero ${\bf P}$ is due to the term $V$ coupling magnetic and ferroelectric orders whose form we discuss below.  Note that this expansion is expressed in the disordered phase, so that all terms in this equation must be invariant under the complete set of symmetry operations of the disordered phase.  In particular, this expression must be invariant under spatial inversion.

 We now discuss the form of the magnetoelectric coupling $V$.  To conserve wavevector this coupling must be at least trilinear,\cite{TRILINEAR} being proportional to one order parameter at wavevector $k$, another at wavevector $(-k)$, and ${\bf P}$ which is associated with zero wavevector. As we have already argued, this coupling is zero with only a single magnetic order parameter  $\sigma_{\rm H}$.  A similar argument made for the LTI variables shows that ferroelectricity can not be induced if we only invoke the single order parameter $\sigma_{\rm L}$. Accordingly, and in analogy with what has been done for the theory of second harmonic generation,\cite{TWO} we posit the following interaction which involves two different symmetry order parameters
\begin{eqnarray}
V &=&  - \sum_{ij \gamma} [ a_{ij\gamma} \sigma_{H;i} (k)\sigma_{L;j}(-k) \nonumber \\ && \ + a_{ij\gamma}^* \sigma_{H;i} (-k) \sigma_{L;j} (k) ] P_\gamma \ .\label{EQ1}
\end{eqnarray}
Using $I\sigma_{X;i}(k)=\sigma_{X;i}(-k)$ and $IP_\gamma =-P_\gamma$, we see that the inversion invariance of $V$ implies that the coefficients $a_{ij\gamma}$ must be purely imaginary.  Using Eq. (\ref{EQ2}) and its analog for the L order parameters we then have
\begin{eqnarray}
V_{LTI} &=&  \sum_\gamma a_\gamma \sigma_H \sigma_L\sin(\phi_H-\phi_L) P_\gamma \ ,\label{EQ3}
\end{eqnarray}
where $a_\gamma$ is a real valued coefficient.  At fourth order in the Landau expansion it can be shown that $\phi_H-\phi_L=\pi/2$, but to have $P_\gamma \not=0$ it is only essential that $\phi_H \not= \phi_L$.(A simple way to understand the effect of the fourth order terms is to recall that Wilson in his original formulation of the renormalization group, used these terms to mimic the constraint of fixed length spins.\cite{KOGUT} Here a fixed length of the spins is most nearly achieved by having the two incommensurate types of order be out of phase by $\pi/2$). We now insert Eq.(\ref{EQ3}) into Eq. (\ref{FEQ}) and minimize with respect to ${\bf P}$. We then find a spontaneous polarization given by
\begin{eqnarray}
P_{\gamma} & \propto & a_\gamma \chi_{\rm el} \sigma_L \sigma_H.\label{EQ4}
\end{eqnarray}
Note that this prediction can be checked by measuring all the quantities which appear in Eq. (7){\it as a function of magnetic field and temperature}. This result on the dependence of the spontaneous polarization on the magnetic order parameter could not be obtained from magnetic symmetry group analysis alone. 

To analyze the consequences of the trilinear coupling in Eq.~(\ref{EQ3}), it is necessary to know how the order parameters transform under the symmetry operations of the crystal.  As stated above, $\sigma_{\rm L}$ is even and $\sigma_{\rm H}$ is odd under a two-fold rotation about the \underline a axis, so for $V$ to be an invariant, ${\bf P}$ must also be odd under this operation.  (This implies that in Eq. (\ref{EQ3}), $\gamma$ can only be $b$ or $c$.)  Furthermore, $\sigma_{\rm L}$ and $\sigma_{\rm H}$ are both even under the \underline a-\underline b glide plane, which restricts $\gamma$ in Eq (\ref{EQ3}) to be in this plane.  Taking account of both symmetries, we see that $a_\gamma$ in Eq. (\ref{EQ3}) can only be nonzero for $\gamma = b$.  This simple argument therefore explains why the spontaneous polarization is confined to the \textbf{b} axis. Furthermore, since a) the polarization depends on the product of $\sigma_L$ and $\sigma_H$ and b) $\sigma_H$ is almost constant in the LTI phase, we expect $P_b$ to be approximately proportional to $\sigma_L$, which is qualitatively what is observed. Note also that we do not expect any coupling between the uniform applied magnetic field, $H_a$, and the uniform polarization, $P_\gamma$, which explains why $P_\gamma$ was observed to be independent of the sign of $H_a$.

In summary: we have shown that the development of ferroelectric order is coincident with an incommensurate magnetic phase in NVO.  Since ferroelectricity occurs only in the phase for which magnetic ordering breaks inversion symmetry, one can reversibly switch the polarization on and off using an external magnetic field.  We develop a Landau-like model to explain how ferroelectricty is induced by incommensurate magnetic ordering.It is expected that this model will explain the behavior of other recently studied multiferroics with incommensurate or long wavelength structure\cite{ferrite,RbFeMoO} and will point the way to microscopic theories for these materials.  In addition, we suggest that the spontaneous development of electric and magnetic long-range order may have been overlooked in simple antiferromagnets in which the lattice structure has inversion symmetry but the magnetic moments only have inversion symmetry relative to a {\it different} origin (Fig 1e). 

We acknowledge support from the LDRD program at LANL and the U.S.-Israel Binational Science Foundation.The NSF supported work at JHU throughDMR-0306940, work at Princeton through DMR-0244254, and work at SPINS through DMR-9986442.

\begin{figure}
\caption{\label{figureA}Crystal and magnetic structures of NVO. \textbf{a}Crystal structure showing
spin-1 Ni$^{2+}$  spine sites in red and cross tie sites in blue.\textbf{b-d} Simplified schematic representation of spin arrangement in the antiferromagnetic HTI, LTI, and CAF phases.\protect\cite{NVO} ``$\bigodot$" indicates the direction of uniform magnetization distributed over spine and cross tie sites in the CAF phase. Only the HTI and CAF phases have inversion symmetry relative to the indicated central lattice point.  Panel \textbf{e} illustrates the center of spatial inversion for the lattice does not coincide with the inversion centre of the magnetic structure.}
\end{figure}

\begin{figure}
\caption{\label{figb}Phase diagram of NVO versus {\em T} and \textbf{H} for \textbf{H}$||$\textbf{a} and \textbf{H}$||$\textbf{c} in panels \textbf{a} and \textbf{b} respectively. The data points indicate anomalies in specific heat ($C$), magnetization ($M$),dielectric permittivity ($\varepsilon$), and electric polarization($P$) traces versus \textbf{H} and {\em T}. Solid lines are guides to the eye. The phases are described in the text and illustrated in Fig. 1\textbf{b-d}}
\end{figure}

\begin{figure}
\caption{\label{figc}Promotion and suppression of electric polarization by applying magnetic fields in NVO. Temperature and magnetic-field dependence of electric polarization along the \textbf{b} axis for \textbf{H} along the \textbf{a} (frames \textbf{a} and \textbf{b})and \textbf{c} (frames \textbf{c} and \textbf{d}) axes.}
\end{figure}

\clearpage
\begin{figure}
\includegraphics[width=7cm,bbllx=120,bblly=140,bburx=560,bbury=700]{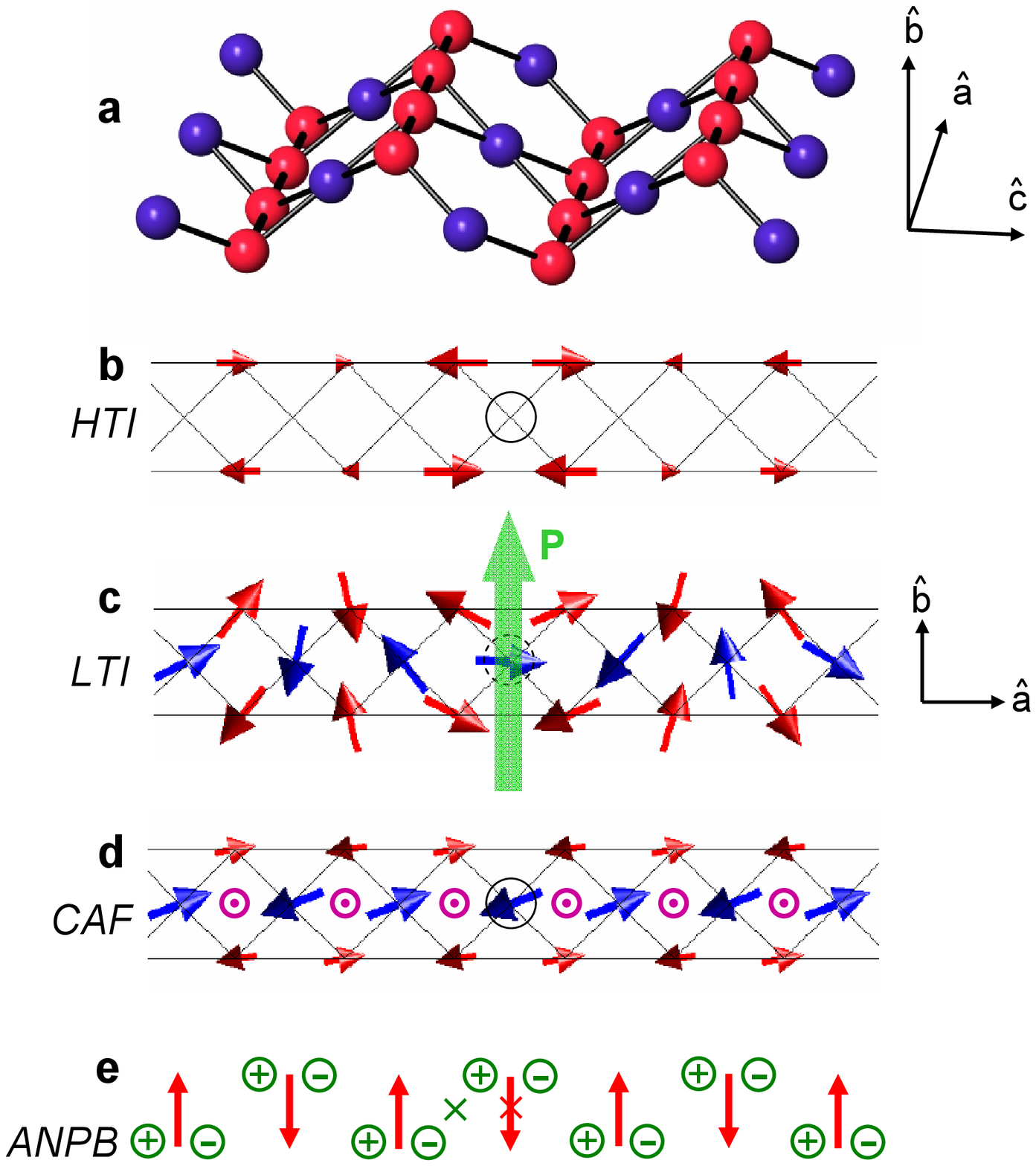}
\end{figure}

\clearpage
\begin{figure}
\includegraphics[width=6.5cm,bbllx=50,bblly=90,bburx=550,bbury=800]{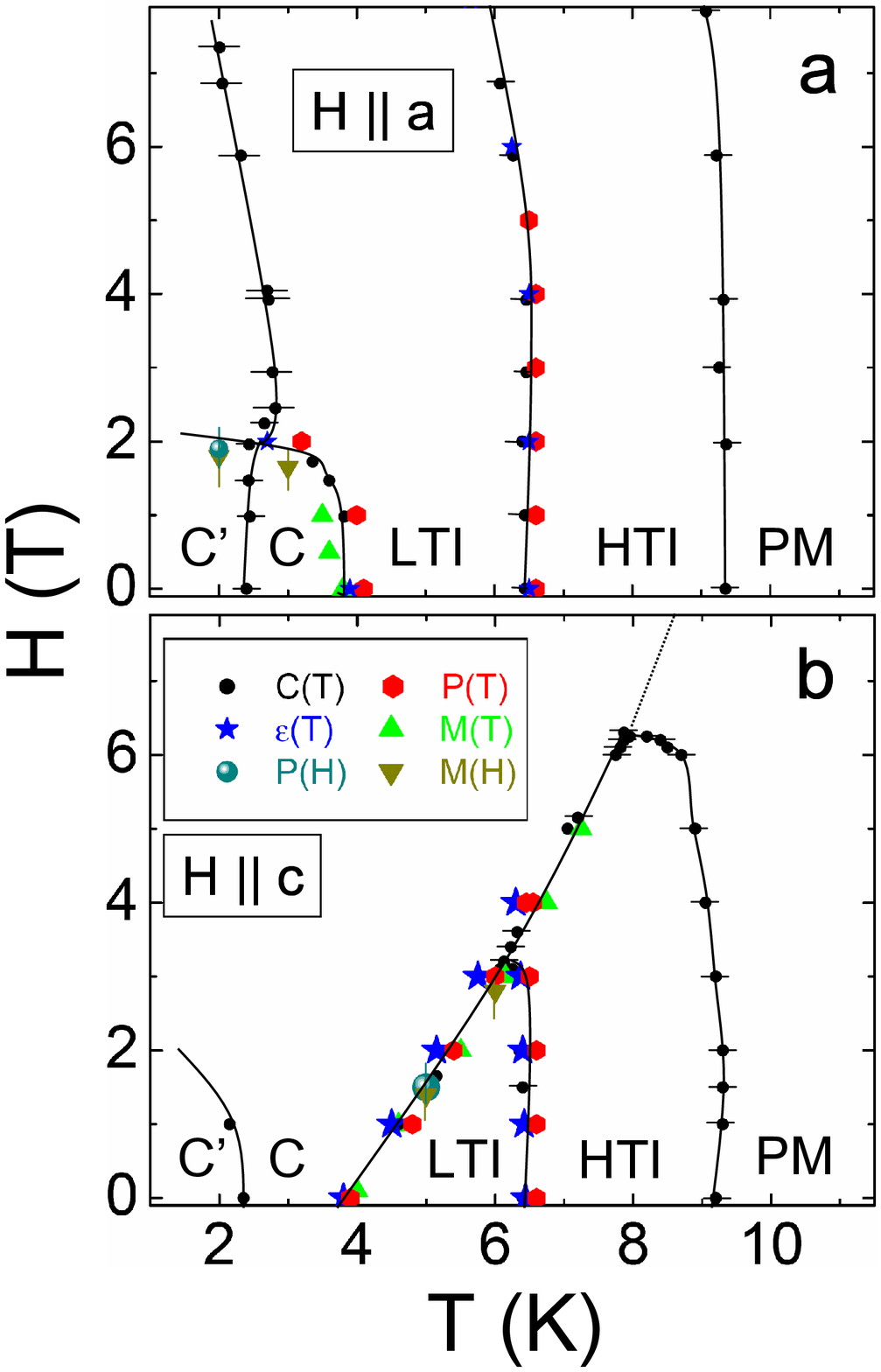}
\end{figure}

\clearpage
\begin{figure}
\includegraphics[width=14cm,angle=-90]{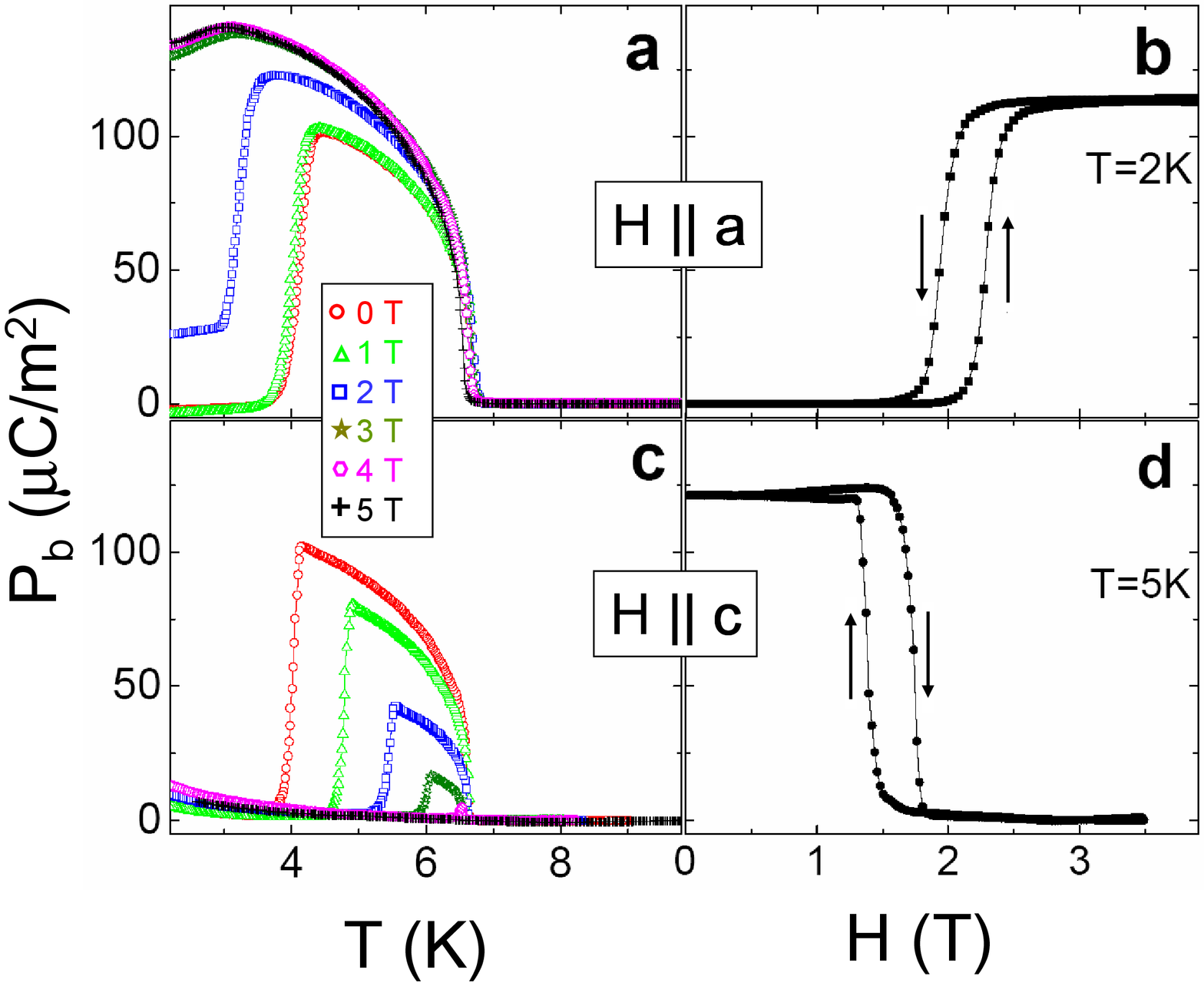}
\end{figure}


\begin{thebibliography} {99}
\bibitem{IC} 
Mathematically ``incommensurate'' means that the modulation vector, $k$,can not be represented as the quotient $n/m$ of two integers.By ``incommensurate'' we mean that $m$ is large enough to betreated as infinite.  Experiment shows that the temperature dependenceof $k$ is smooth at the level of about 1/2 \%, which implies that$m\geq 50$.
\bibitem{TRILINEAR} 
The coupling described in Eq. (5) is also sometimes referred toas``bilinear" in the magnetoelectric literature.
\end{thebibliography}
\end{document}